# The Missing Link of Sulfur Chemistry in TMC-1: The Detection of $c$-C$_3$H$_2$S from the GOTHAM Survey

Anthony J. Remijan,[1] P. Bryan Changala,[2] Ci Xue,[3] Elsa Q.H. Yuan,[4] Miya Duffy,[3] Haley N. Scolati,[5] Christopher N. Shingledecker,[6] Thomas H. Speak,[4] Ilsa R. Cooke,[4] Ryan Loomis,[1] Andrew M. Burkhardt,[7] Zachary T.P. Fried,[3] Gabi Wenzel,[3,2] Andrew Lipnicky,[1] Michael C. McCarthy,[2] and Brett A. McGuire[3,1]

[1]*National Radio Astronomy Observatory, Charlottesville, VA 22903, USA*
[2]*Center for Astrophysics | Harvard & Smithsonian, Cambridge, MA 02138, USA*
[3]*Department of Chemistry, Massachusetts Institute of Technology, Cambridge, MA 02139, USA*
[4]*Department of Chemistry, University of British Columbia, 2036 Main Mall, Vancouver, BC V6T 1Z1, Canada*
[5]*Department of Chemistry, University of Virginia, Charlottesville, VA 22904, USA*[a]
[6]*Department of Chemistry, Virginia Military Institute, Lexington, VA 24450*
[7]*Department of Earth, Environment, and Physics, Worcester State University, Worcester, MA 01602, USA*

## ABSTRACT

We present the spectroscopic characterization of cyclopropenethione ($c$-C$_3$H$_2$S) in the laboratory and detect it in space using the Green Bank Telescope (GBT) Observations of TMC-1: Hunting Aromatic Molecules (GOTHAM) survey. The detection of this molecule – the missing link in understanding the C$_3$H$_2$S isomeric family in TMC-1 – completes the detection of all 3 low-energy isomers of C$_3$H$_2$S as both CH$_2$CCS and HCCCHS have been previously detected in this source. The total column density of this molecule ($N_T$ of $5.72^{+2.65}_{-1.61} \times 10^{10}$ cm$^{-2}$ at an excitation temperature of $4.7^{+1.3}_{-1.6}$ K) is smaller than both CH$_2$CCS and HCCCHS and follows nicely the relative dipole principle (RDP), a kinetic rule-of-thumb for predicting isomer abundances which suggests that, all other chemistry among a family of isomers being the same, the member with the smallest dipole ($\mu$) should be the most abundant. The RDP now holds for the astronomical abundance ratios of both the S-bearing and O-bearing counterparts observed in TMC-1; however, CH$_2$CCO continues to elude detection in any astronomical source.

## 1. INTRODUCTION

Sulfur is one of the more abundant elements in the interstellar medium, rounding out the top 10 most abundant molecules with an abundance of ∼ 10$^{-5}$ with respect to hydrogen (Asplund et al. 2005; Howk et al. 2006). Sulfur-bearing molecules offer valuable diagnostics into the physical and chemical properties of astronomical regions (see e.g. Charnley 1997; Stäuber et al. 2005; Esplugues et al. 2014; Gorai et al. 2024, and references therein). For instance, small diatomic molecules like CS are commonly used as "dense gas tracers" because they are typically found in regions of higher extinction – where gas and dust densities are elevated – and are less readily detected in lower-density regions dominated by CO (see e.g. Zhou et al. 1993). Similarly, SO and SO$_2$ are often considered "shock tracers," as they are frequently associated with protostellar jets of material ejected by newly forming stars (Artur de la Villarmois et al. 2023; Esplugues et al. 2014).

However, it remains unclear what process (or processes) drive the production of these sulfur-bearing species in the wide variety of astronomical environments. Are physical conditions alone sufficient to account for their selective detection, or is it a combination of both physical and chemical factors? For example, CS is now consistently observed in both diffuse and dense gas (Gerin et al. 2024), and SO$_2$ can be found in the cold molecular envelopes of hot core regions just as readily as in shock-heated gas (e.g. Tychoniec et al. 2021). From a chemical perspective, it is also intriguing to explore how the abundances and formation pathways of S-containing molecules compare to their oxygen-bearing counterparts as oxygen and sulfur exhibit similar electron configurations, with six valence electrons that enable them to form two covalent bonds in neutral molecules.

---

[a] Currently a Grote Reber Fellow of the National Radio Astronomy Observatory

Effectively probing these chemical formation and destruction pathways requires controlled tests where the effects of variables not of interest are minimized. To this end, observations of families of molecular isomers are particularly powerful (see, e.g. Xue et al. 2019, and references therein). Recent interest has been drawn to the C$_3$H$_2$O isomeric family of molecules in order to test the relative influence of kinetic versus thermodynamic control (Loison et al. 2016). Notably, the most stable isomer, propadienone (CH$_2$CCO) remains undetected in astronomical sources while its higher-energy structural isomers, propynal (HCCCHO) and cyclopropenone ($c$-C$_3$H$_2$O) have been known and studied for decades (Irvine et al. 1988; Hollis et al. 2006; Loomis et al. 2015; Loison et al. 2016; Shingledecker et al. 2019). This raises important questions: Do we fully understand the chemical formation pathways responsible for this discrepancy? Moreover, why has HCCCHO exclusively been detected in extremely cold regions (temperatures below 10 K), while evading detection in both high- and low-mass hot molecular cores?

Even more curious is the recent detection of the S-analog of this species towards TMC-1 – propadienethione (CH$_2$CCS) (Cernicharo et al. 2021b). High-sensitivity spectral line observations of TMC-1 were performed using the Yebes 40-m telescope as part of the Q-band Ultrasensitive Inspection Journey to the Obscure Tmc-1 Environment (QUIJOTE) survey (Cernicharo, José et al. 2022). From these observations, CH$_2$CCS was identified with a total column density of $3.7 \pm 0.4 \times 10^{11}$ cm$^{-2}$ and a derived rotational temperature of $10 \pm 1$ K. The ortho/para ratio was also measured to be $4.6 \pm 0.8$, consistent with the canonical value of 3:1 expected from the statistical spin degeneracies (Yocum et al. 2023). The HCCCHS isomer was previously detected by Cernicharo et al. (2021a) also as part of the QUIJOTE survey where a total column density of $3.2 \pm 0.4 \times 10^{11}$ cm$^{-2}$ was determined assuming a rotational temperature of 5 K. Thus, the relative column density ratio between these two isomers approaches $\sim 1$, significantly different than the upper limit to the column density ratio of $< 3\%$ between CH$_2$CCO and HCCCHO (Loison et al. 2016).

Although the sulfur depletion problem remains a puzzle of the chemistry in dense clouds (Ruffle et al. 1999; Vidal et al. 2017), the claim by Cernicharo et al. (2021b) that the dark cloud region TMC-1 is a "chemical factory for S-bearing molecules" is well justified because many additional S-bearing species have been detected towards TMC-1 that have not yet been identified in any other astronomical environment. In addition, the column densities of some S-bearing species are higher compared to their O-bearing analogs (see Cernicharo et al. 2021a, and references therein). Yet, why are some of the S-species so prevalent compared to the O-species in this source and what makes the environment of TMC-1 so unique to form the complex S-bearing species? These questions underscore the need to refine our understanding of the relative contribution between physical and chemical processes in determining what molecules are observed in astronomical sources.

To this end, cyclopropenethione ($c$-C$_3$H$_2$S) is a missing link in understanding the C$_3$H$_2$S isomeric family in TMC-1. Spectroscopic work on the C$_3$H$_2$S isomers, specifically HCCCHS and CH$_2$CCS, was first published in the 1980's (Brown et al. 1982, 1988), yet no experimental spectroscopy of $c$-C$_3$H$_2$S has been reported until the present work. Only recently have theoretical studies aimed to characterize its geometry and other spectroscopic properties, such as the rotational partition function, been conducted (Song et al. 2022).

In this paper, we present the spectroscopic characterization of $c$-C$_3$H$_2$S in both the laboratory and in space and make the astronomical detection using the Green Bank Telescope (GBT) Observations of TMC-1: Hunting Aromatic Molecules (GOTHAM) survey. Section 2 describes the theoretical predictions used to determine the initial rotational constants of this molecule and the laboratory observations that led to its astronomical detection. Section 3 presents the observational details of the GOTHAM large program. Section 4 presents the results of the astronomical detection of $c$-C$_3$H$_2$S; section 5 discusses the detection of $c$-C$_3$H$_2$S in the context of the detections of the other S-bearing species detected towards TMC-1 and their O-bearing analogs, and finally section 6 summarizes our conclusions.

## 2. SPECTROSCOPIC ANALYSIS

The laboratory search for the microwave spectrum of $c$-C$_3$H$_2$S was guided by high-level quantum chemical predictions of its spectroscopic parameters calculated with the CFOUR program package (Matthews et al. 2020; Stanton et al. 2020). The equilibrium geometry was optimized with coupled cluster theory including single, double, and perturbative triple excitations [CCSD(T)] (Bartlett et al. 1990; Raghavachari et al. 1989) using analytic gradient methods (Lee

& Rendell 1991) and the Dunning correlation-consistent polarized core-valence basis sets (cc-pCVXZ, with $X$ = D, T, Q, 5) (Dunning 1989; Woon & Dunning 1995; Peterson & Dunning 2002), correlating all electrons except the sulfur 1s inner-core electrons. The optimized structural parameters, equilibrium rotational constants, and electric dipole moments are summarized in Table 1. Cubic force fields computed with the cc-pCVDZ and cc-pCVTZ basis sets and finite differences of analytic gradients were used to derive the harmonic centrifugal distortion parameters and anharmonic vibrational zero-point corrections with second-order vibrational perturbation theory (VPT2) (Mills 1972) (see Table 1). The best-estimate ground-state rotational constants, which combine the CCSD(T)/cc-pCV5Z equilibrium geometry with the CCSD(T)/cc-pCVTZ vibrational corrections, are $A_0$ = 32128.77, $B_0$ = 4081.75, and $C_0$ = 3618.26 MHz.

**Table 1.** The theoretical CCSD(T) structural parameters and spectroscopic constants of $c$-$C_3H_2{}^{32}S$.

| Parameter | cc-pCVDZ | cc-pCVTZ | cc-pCVQZ | cc-pCV5Z |
|---|---|---|---|---|
| $r(C_1S)$ (Å) | 1.634365 | 1.619490 | 1.616091 | 1.615162 |
| $r(C_1C_2)$ (Å) | 1.447876 | 1.426873 | 1.421809 | 1.420597 |
| $r(C_2H)$ (Å) | 1.092800 | 1.077818 | 1.076477 | 1.076160 |
| $\theta(SC_1C_2)$ (°) | 152.124 | 152.099 | 152.069 | 152.062 |
| $\theta(HC_2C_1)$ (°) | 150.540 | 150.175 | 150.096 | 150.063 |
| $\mu_e$ (D)[a] | 4.146 | 4.427 | 4.549 | 4.572 |
| $A_e$ (MHz)[b] | 31313.58 | 32137.66 | 32273.07 | 32304.46 |
| $B_e$ (MHz)[b] | 3971.54 | 4066.36 | 4089.65 | 4095.64 |
| $C_e$ (MHz)[b] | 3524.52 | 3609.64 | 3629.70 | 3634.81 |
| $A_e - A_0$ (MHz)[c] | 172.74 | 175.69 | — | — |
| $B_e - B_0$ (MHz)[c] | 13.70 | 13.89 | — | — |
| $C_e - C_0$ (MHz)[c] | 16.23 | 16.55 | — | — |
| $D_J$ (kHz)[d] | 0.466 | 0.495 | — | — |
| $D_K$ (kHz)[d] | 55.094 | 58.203 | — | — |
| $D_{JK}$ (kHz)[d] | 16.833 | 17.679 | — | — |
| $d_1$ (Hz)[d] | −62.800 | −66.162 | — | — |
| $d_2$ (Hz)[d] | −20.569 | −21.459 | — | — |

[a] The equilibrium electric dipole moment.
[b] The equilibrium rotational constants.
[c] The VPT2 ground-state vibrational corrections.
[d] The S-reduced ($I^r$) quartic centrifugal distortion parameters.

The microwave spectrum of $c$-$C_3H_2S$ was measured with two cavity Fourier transform microwave (FTMW) spectrometers operating over 5–26 GHz and 5–40 GHz, respectively, coupled to supersonic electric discharge expansion sources (Crabtree et al. 2016; Grabow et al. 2005), an approach previously used in our laboratory to produce a diverse array of other reactive S-bearing molecules (McCarthy et al. 2003; Gordon et al. 2001, 2002; Lee et al. 2019; McGuire et al. 2018). Dilute mixtures of a hydrocarbon precursor (0.1% HCCH) and a sulfur source (either 0.1% $CS_2$ or $H_2S$) were prepared in neon and expanded through a pulsed solenoid valve into a large vacuum chamber containing a Fabry-Pérot microwave cavity. Before expanding into the cavity, the gas mixture passed through two copper ring electrodes spaced by ~1 cm and held at an electric potential difference of 0.8–1.2 kV, which generated an electric discharge (30 mA peak current) during the ~300–600 μs gas pulse. The mixture supersonically expanded along the axis of the microwave cavity and cooled to an internal rotational temperature of approximately 2 K.

An initial laboratory search for $c$-$C_3H_2S$ was performed using discharge expansion conditions optimized for the production of propadienethione, $\ell$-$H_2C_3S$, a lower-energy chain isomer of $c$-$C_3H_2S$ (Brown et al. 1988; McCarthy et al. 2021). We first targeted the $J_{K_a,K_c}$ = $2_{1,2}$−$1_{1,1}$ transition of $c$-$C_3H_2S$, which was predicted to lie near 14936 MHz based on the best-estimate theoretical rotational constants. Six molecular features were detected within a ±30 MHz window centered around the predicted transition frequency, but only one (at 14931.538(2) MHz) could possibly be assigned to

**Table 2.** The laboratory rest frequencies of $c$-C$_3$H$_2$S.

| Transition $J'\ K_a'\ K_c' - J''\ K_a''\ K_c''$ | Rest Frequency (MHz)[a] $c$-C$_3$H$_2{}^{32}$S | $c$-C$_3$H$_2{}^{34}$S |
|---|---|---|
| 1 0 1 – 0 0 0 | 7697.4252 | 7491.2114 |
| 2 1 2 – 1 1 1 | 14931.5383 | 14543.5090 |
| 2 0 2 – 1 0 1 | 15389.1452 | 14977.3261 |
| 2 1 1 – 1 1 0 | 15857.9893 | 15421.1810 |
| 3 1 3 – 2 1 2 | 22393.7444 | 21812.0748 |
| 3 0 3 – 2 0 2 | 23069.4627 | 22453.2381 |
| 3 2 2 – 2 2 1 | 23091.7924 | |
| 3 2 1 – 2 2 0 | 23114.5446 | |
| 3 1 2 – 2 1 1 | 23783.3699 | 23128.5379 |
| 4 1 4 – 3 1 3 | 29851.7385 | 29076.8676 |
| 4 0 4 – 3 0 3 | 30732.7075 | 29913.8729 |
| 4 2 3 – 3 2 2 | 30784.5657 | |
| 4 2 2 – 3 2 1 | 30841.4213 | |
| 4 1 3 – 3 1 2 | 31704.3514 | 30831.9675 |
| 5 1 5 – 4 1 4 | 37304.2121 | |
| 5 0 5 – 4 0 4 | 38373.2945 | |
| 5 1 4 – 4 1 3 | 39619.3638 | |

[a] All frequencies have an uncertainty of 2 kHz.

$c$-C$_3$H$_2$S: this line was observed only when the discharge was enabled; was not observed when HCCH, CS$_2$, or H$_2$S was removed; and was unaffected by an applied magnetic field, indicating a closed-shell species. A candidate for the $2_{1,1} - 1_{1,0}$ transition predicted to lie near 15863 MHz was subsequently detected at 15857.989(2) MHz and passed the same series of tests. After a preliminary fit of the $B$ and $C$ rotational constants based on these 2 tentative assignments, 15 additional transitions were detected, totaling 17 lines between 7 and 40 GHz. The measured rest frequencies were reproduced to within the experimental uncertainty (2 kHz) by adjusting the three rotational constants ($A$, $B$, and $C$) and two quartic centrifugal distortion (CD) constants ($D_J$ and $D_{JK}$), leaving the remaining CD constants fixed to their CCSD(T)/cc-pCVTZ theoretical values. The laboratory rest frequencies and derived spectroscopic constants of $c$-C$_3$H$_2$S are summarized in Tables 2 and 3, respectively.

The experimentally derived rotational constants of $c$-C$_3$H$_2$S differ from the best-estimate theoretical predictions by less than 0.04%. Taken together with the experimental tests described above, the newly detected molecule would almost certainly be $c$-C$_3$H$_2$S, but definitive confirmation of the assignment was ultimately secured by the observation of the $^{34}$S isotopic species. Guided by the experimentally derived rotational constants of the parent isotopologue corrected by isotope shifts derived from the CCSD(T)/cc-pCV5Z equilibrium structure, we detected 10 rotational transitions of $c$-C$_3$H$_2\ {}^{34}$S between 7 and 31 GHz (Table 2). Their intensities relative to the corresponding transitions of the parent species were consistent with the natural abundance of $^{34}$S (4%). The best-fit rotational constants of $c$-C$_3$H$_2\ {}^{34}$S (Table 3) differ from those of the parent species by $\Delta B = -115.3$ MHz and $\Delta C = -90.9$ MHz, which agree with the theoretical equilibrium shifts ($-115.8$ MHz and $-91.5$ MHz, respectively) to better than 1%. The high sensitivity of these isotopic shifts to the molecular structure leaves no doubt as to the laboratory identification of $c$-C$_3$H$_2$S. The full measured line list along with the corresponding input and output files from SPFIT/SPCAT are provided as Supplemental Information. The calculated rotational partition function values for $c$-C$_3$H$_2$S at distinct temperatures from 1 – 500 K are given in Table 4.

### 3. OBSERVATIONS

Data for this study were gathered as part of the GOTHAM survey, a comprehensive program utilizing the 100-m GBT. This analysis uses the fifth data reduction (DRV) from the GOTHAM survey, focusing on the cyanopolyyne peak (CP) in TMC-1, centered at $\alpha_{J2000} = 04^{\rm h}41^{\rm m}42.5^{\rm s}$, $\delta_{J2000} = +25°41'26.8''$. Detailed information about the data reduction procedures will be available in Xue et al. (in preparation), while the observing strategy is described in McGuire et al. (2020). In summary, the GOTHAM survey spectra cover the C-, X-, Ku-, K-, and Ka-receiver bands, providing nearly continuous coverage from 3.9 to 11.6 GHz, 12.7 to 15.6 GHz, and 18.0 to 36.4 GHz, totaling 29 GHz of bandwidth. All observations were made with a consistent frequency resolution of 1.4 kHz (equivalent to $0.05 - 0.01$ km/s in velocity) and a root-mean-square (RMS) noise ranging from approximately 2 to 20 mK across most of the frequency range. The RMS noise varies across frequency ranges mainly due to different total integration times and zenith opacity corrections. Data reduction involved removing radio frequency interference (RFI) and artifacts, fitting the baseline continuum, and calibrating the flux density scale using the internal noise diodes. The uncertainty in flux calibration is estimated to be $10 - 20\%$ so a 20% systematic uncertainty is included in the statistical analysis detailed below.

**Table 3.** Rotational constants of $c$-$C_3H_2S$.[a]

| Parameter | $c$-$C_3H_2{}^{32}S$ | $c$-$C_3H_2{}^{34}S$ | Theory[b] |
|---|---|---|---|
| $A$ (MHz) | 32121.0(6) | 32114.8(8) | 32128.77 |
| $B$ (MHz) | 4080.3272(3) | 3965.0286(3) | 4081.75 |
| $C$ (MHz) | 3617.1000(3) | 3526.1901(3) | 3618.26 |
| $D_J$ (kHz) | 0.519(6) | [0.519][d] | |
| $D_K$ (kHz) | [58.203][c] | [58.203][d] | |
| $D_{JK}$ (kHz) | 18.34(6) | [18.34][d] | |
| $d_1$ (Hz) | [−66.162][c] | [−66.162][d] | |
| $d_2$ (Hz) | [−21.459][c] | [−21.459][d] | |
| Transitions in fit | 17 | 10 | |
| $\sigma_{fit}$ (kHz) | 1.7 | 3.4 | |

[a] The 1-$\sigma$ uncertainties are shown in parentheses in units of the last digit.
[b] Best-estimate theoretical ground-state rotational constants of $c$-$C_3H_2{}^{32}S$ derived from the CCSD(T)/cc-pCV5Z equilibrium geometry with the CCSD(T)/cc-pCVTZ vibrational corrections (Table 1).
[c] Fixed to theoretical values (Table 1).
[d] Fixed to the values of the $^{32}S$ species.

**Table 4.** Partition function for $c$-$C_3H_2S$ as determined by SPCAT and used in the MCMC analysis.

| Temperature [K] | Partition function |
|---|---|
| 1.0 | 13.1482 |
| 1.5 | 26.8913 |
| 2.0 | 43.2537 |
| 2.5 | 61.5061 |
| 3.0 | 81.3666 |
| 3.5 | 102.7169 |
| 4.0 | 125.4909 |
| 4.5 | 149.6359 |
| 5.0 | 175.1010 |
| 5.5 | 201.8358 |
| 6.0 | 229.7918 |
| 6.5 | 258.9228 |
| 7.0 | 289.1862 |
| 7.5 | 320.5426 |
| 8.0 | 352.9561 |
| 8.5 | 386.3935 |
| 9.375 | 447.2832 |
| 18.75 | 1260.3705 |
| 37.5 | 3500.4983 |
| 75.0 | 9127.3074 |
| 150.0 | 21741.8704 |
| 225.0 | 34841.8597 |
| 275.0 | 43667.8052 |
| 300.0 | 48095.9986 |
| 400.0 | 65868.1649 |
| 500.0 | 83682.1758 |

## 4. ASTRONOMICAL ANALYSIS

Best fit parameters for source size (″), column density ($N_T$), excitation temperature ($T_{ex}$), line width ($\Delta V$), and velocity with respect to the local standard of rest, ($v_{LSR}$), were determined using a Markov chain Monte Carlo (MCMC) fit to the data as described in (Loomis et al. 2021). For the analysis of $c$-$C_3H_2S$, a total of 52 lines are used over the entire frequency range covered by the GOTHAM survey. We follow the model assumption of four separate components with distinct systematic velocities, as revealed by recent high spectral resolution observations (Dobashi et al. 2018; Xue et al. 2020). Uniform priors were applied to the source size and $N_T$, while Gaussian priors were applied for $T_{ex}$, $v_{LSR}$, and $\Delta V$ (Table 5). Because of the limited range of energy levels of $c$-$C_3H_2S$ transitions detected by the GOTHAM data, the prior for its $T_{ex}$ was modeled as a Gaussian distribution centered at 4.2 K. This was derived from the posterior distribution of the well-characterized oxygen analog $c$-$C_3H_2O$ (Table 9 in Appendix A) which has bright individual lines and was fitted first with uninformative uniform distributed priors, except for $v_{LSR}$, which was constrained by previous GOTHAM analyses (Loomis et al. 2021). The $c$-$C_3H_2O$ posteriors were therefore used as priors for $c$-$C_3H_2S$ for $T_{ex}$, $v_{LSR}$, and $\Delta V$. Posterior probability distributions and covariances for each physical parameter were derived for four distinct velocity components over 30,000 iterations using 100 walkers.

**Table 5.** Priors used for MCMC analysis of $c$-$C_3H_2S$ observations towards TMC-1. $U\{a, b\}$ indicates a uniform distribution with minimum, $a$, and maximum, $b$, while $N(\mu, \sigma^2)$ indicates a Gaussian distribution with mean, $\mu$, and variance, $\sigma^2$. The Gaussian priors for $T_{ex}$, $v_{LSR}$, and $\Delta V$ are informed by the posterior distribution of $c$-$C_3H_2O$.

| Component No. | $v_{LSR}$ [km s$^{-1}$] | Size [″] | $\log_{10}(N_T)$ [cm$^{-2}$] | $T_{ex}$ [K] | $\Delta V$ [km s$^{-1}$] |
|---|---|---|---|---|---|
| 1 | $N(5.624, 0.01)$ | $U\{0, 300\}$ | | | |
| 2 | $N(5.781, 0.01)$ | $U\{0, 300\}$ | $U\{9.0, 12.0\}$ | $N\{4.2, 3.0\}$ | $N\{0.158, 0.05\}$ |
| 3 | $N(5.915, 0.01)$ | $U\{0, 300\}$ | | | |
| 4 | $N(6.018, 0.01)$ | $U\{0, 300\}$ | | | |

To evaluate the statistical evidence that our emission model for these molecules aligns with the data, we followed the procedures detailed in Loomis et al. (2021), performing a spectral stacking and matched filtering analysis for $c$-$C_3H_2S$. Briefly, this involved computing a weighted average of the observational spectra in velocity space, centered on the observed frequency of each spectral line of the target molecule. The weights were determined by the relative intensity of the expected emission (based on spectral predictions using the MCMC-derived parameters) and the local RMS noise of the observations. Due to the weak expected intensities for $c$-$C_3H_2S$, any observational windows showing emission stronger than $>5\sigma$ were excluded from the analysis to prevent interference from other species.

Simulated spectra of the molecular emission were also generated using the same MCMC-derived parameters and stacked using identical weights. This simulation served as a matched filter, passed through the observational data to produce an impulse response function. The response function quantified the statistical evidence for our emission model – and the derived molecular parameters – being consistent with the observations. The appendices of McGuire et al. (2021) complement the methodology described in Loomis et al. (2021) with an extensive analysis of its robustness, addressing the improbability of spurious signals and the minimal impact of non-thermal noise terms on the results.

Based on this analysis, we report the first interstellar detection of $c$-$C_3H_2S$ toward TMC-1. Figure 1 presents the stacked observational data and the stacked MCMC model on the left panel, and the matched filter response on the right panel. The statistical significance of this detection is $5.4\sigma$ as shown in the right panel of Figure 1, confirming the presence of this molecule. The posterior probability distribution for the MCMC fit is shown in Figure 5 and summarized in Table 6, suggesting a total $N_T$ of $5.72^{+2.65}_{-1.61} \times 10^{10}$ cm$^{-2}$ for $c$-$C_3H_2S$ at an excitation temperature of $4.7^{+1.3}_{-1.1}$ K. Uncertainties, corresponding to the 16th and 84th percentiles ($\pm 1\sigma$ for a Gaussian posterior distribution), are reported for each parameter.

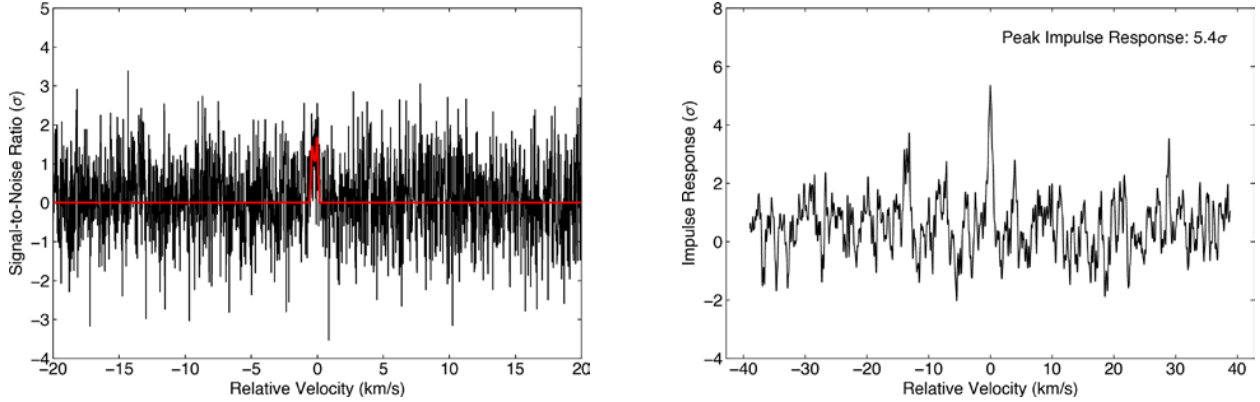

**Figure 1.** Velocity stacked and matched filter spectra of $c$-$C_3H_2S$. The intensity scales are the signal-to-noise ratios (SNR) of the response functions when centered at a given velocity. The "zero" velocity corresponds to the channel with the highest intensity to account for blended spectroscopic transitions and variations in velocity component source sizes. (*Left*) The stacked spectra from the GOTHAM DRV data are displayed in black, overlaid with the expected line profile in red from our MCMC fit to the data. The SNR is on a per-channel basis. (*Right*) Matched filter response obtained from cross-correlating the simulated and observed velocity stacks in the left panel; value annotated corresponds to the peak impulse response of the matched filter.

In addition, for the purposes of the comparison between isomeric members, we have therefore additionally performed MCMC fittings for the other isomeric members of $H_2C_3S$ and $H_2C_3O$ with a spatial-separate four-component model, with results presented in detail in Appendix A. In addition to $c$-$C_3H_2S$, $c$-$C_3H_2O$ and $CH_2CCS$ are distinctly detected whereas no clear detection of $CH_2CCO$ nor HCCCHS is made. A total $N_T$ of $6.64^{+2.02}_{-0.71} \times 10^{11}$ cm$^{-2}$ is determined for $c$-$C_3H_2O$ and $5.80^{+0.85}_{-0.61} \times 10^{11}$ cm$^{-2}$ for $CH_2CCS$. And an upper limit of $< 3.76 \times 10^{11}$ cm$^{-2}$ is determined for $CH_2CCO$ and $< 9.51 \times 10^{11}$ cm$^{-2}$ for HCCCHS.

**Table 6.** Summary statistics of the marginalized $c$-$C_3H_2S$ posterior.

| $v_{LSR}$ (km s$^{-1}$) | Size (″) | $N_T$ ($10^{10}$cm$^{-2}$) | $T_{ex}$ (K) | $\Delta V$ (km s$^{-1}$) |
|---|---|---|---|---|
| $5.621^{+0.01}_{-0.01}$ | $148^{+103}_{-101}$ | $2.11^{+1.67}_{-0.86}$ | | |
| $5.782^{+0.01}_{-0.01}$ | $146^{+105}_{-102}$ | $1.04^{+1.13}_{-0.75}$ | $4.7^{+1.3}_{-1.1}$ | $0.171^{+0.024}_{-0.024}$ |
| $5.914^{+0.01}_{-0.01}$ | $151^{+102}_{-104}$ | $1.03^{+1.25}_{-0.75}$ | | |
| $6.035^{+0.01}_{-0.01}$ | $158^{+95}_{-99}$ | $1.53^{+1.18}_{-0.88}$ | | |
| $N_T$(Total): $5.72^{+2.65}_{-1.61} \times 10^{10}$ cm$^{-2}$ | | | | |

## 5. DISCUSSION

The relative abundances of isomers in the interstellar medium (ISM) can provide critical astrochemical clues as to the underlying chemical reactions at play in astrophysical environments, and the detection of $c$-$C_3H_2S$ reported here is no exception. As noted, the analogous $C_3H_2O$ family of isomers — namely, cyclopropenone ($c$-$C_3H_2O$), propynal (HCCCHO), and propadienone ($CH_2CCO$) — have been the subject of a number of studies (Gorai et al. 2024; Loison et al. 2016), with many focused on the curious absence of propadienone, the most thermodynamically stable of the three (Loomis et al. 2015). As was discussed in Loomis et al. (2015), and later by Shingledecker et al. (2019), the consistent non-detection of propadienone underscores the importance of kinetic factors in setting the abundances of astrochemical species in the ISM. In particular, Shingledecker et al. (2019) found from ab initio calculations that propadienone, alone among the three $C_3H_2O$ isomers, was highly reactive with atomic hydrogen, which would thereby lower its abundance relative to the other two. Along those lines, Shingledecker et al. (2020) proposed the relative dipole principle (RDP), a kinetic rule-of-thumb for predicting isomer abundances. Briefly, the RDP predicts that, all other chemistry among a family of isomers being the same, the member with the smallest dipole ($\mu$) should be the most abundant, and vice-versa, since the rate of destruction via reaction with gas-phase ions is proportional to $\mu$.

**Table 7.** Summary of the $C_3H_2O$ and $C_3H_2S$ detections towards TMC-1.

| Molecule | Dipole[a] (D) | ZPBE[b] (kJ mol$^{-1}$) | $N_T$ ($10^{10}$ cm$^{-2}$) | Reference |
|---|---|---|---|---|
| $CH_2CCO$ | 2.16 | 0 | <37.6 | This work |
| HCCCHO | 2.36 | 22.9 | $728^{+408}_{-194}$ | Remijan et al. (2024) |
| $c$-$C_3H_2O$ | 4.39 | 26.3 | $66.4^{+20.2}_{-7.1}$ | This work |
| $CH_2CCS$ | 2.06 | 0 | $58.0^{+8.5}_{-6.1}$ | This work |
| HCCCHS | 1.88 | 1.5 | 32.0 | Cernicharo et al. (2021a) |
| $c$-$C_3H_2S$ | 4.57 | 11.7 | $5.72^{+2.65}_{-1.61}$ | This work |

[a] Value for the a-type dipole moment only.
[b] Zero Point Bonding Energy.

Turning our attention once more to the $C_3H_2S$ family, we will first examine the relative abundances predicted by the RDP, and then examine the astrochemical implications of deviations from it. To guide this discussion, the *a*-type dipole moments, zero point bonding energy and total reported column densities of the $C_3H_2O$ and $C_3H_2S$ family of isomers are listed in Table 7.

We begin by assuming dipole values of 4.57 D for $c$-$C_3H_2S$ (Table 6), as well as of 2.06 D (Cernicharo et al. 2021b) and 1.88 D (Cernicharo et al. 2021a) for $CH_2CCS$ and HCCCHS, respectively. The RDP prediction is thus that $c$-$C_3H_2S$ should be the least abundant, while HCCCHS should be slightly more abundant than $CH_2CCS$. As shown in Table 7, the derived column density for $c$-$C_3H_2S$ is indeed the lowest relative to the other 2 isomers. The column densities of the other two isomers are also generally in line with the expected RDP trend, taking into account their reported uncertainties from this work and from the column density for HCCCHS reported by Cernicharo et al. (2021a). This agreement between the observational relative abundances and those predicted by the RDP indicates that, though the formation of $c$-$C_3H_2S$ likely occurs via a different route than the other two, its overall rate of formation is likely similar to $CH_2CCS$ and HCCCHS (see Subsection 5.1 below).

Based on the results of this analysis, two surprising outcomes emerge. First, there is a notable discrepancy in the relative column densities measured between the lowest and highest energy isomers of the O- and S-bearing species. Specifically, for the O-bearing molecules, the cyclic isomer exhibits a significantly *higher* column density than the linear isomer. In contrast, the S-bearing species display the opposite trend, with the cyclic isomer present at much *lower* column densities than the linear counterpart. Second, this observation highlights a striking chemical implication: as shown in Shingledecker et al. (2019), the linear form of the $C_3H_2O$ isomer is highly reactive with atomic hydrogen in this region and in a manner which must be fundamentally different from its S-bearing analog. The reasons behind this behavior and the factors responsible for this remarkable difference in reactivity remain unknown.

### 5.1. Ab-initio evaluation of a potential formation route to c-$C_3H_2S$

Though a thorough study of the formation of c-$C_3H_2S$ is beyond the scope of this work, by analogy to the c-$C_3H_2O$ formation pathway

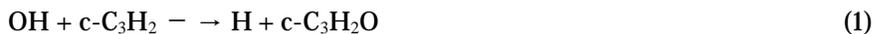

$$OH + c\text{-}C_3H_2 \longrightarrow H + c\text{-}C_3H_2O \tag{1}$$

from Loison et al. (2017), we posit that a plausible formation route for c-$C_3H_2S$ could be

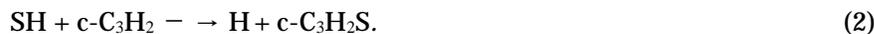

$$SH + c\text{-}C_3H_2 \longrightarrow H + c\text{-}C_3H_2S. \tag{2}$$

In light of this current work, we conducted an *ab-initio* study to assess the feasibility of reaction (2).

Computational work to probe potential routes to the formation of the $C_3H_2S$ family ($c$-$C_3H_2S$, $H_2CCCS$, HCCCHS) from the reaction of SH and $c$-$C_3H_2$ was carried out. This work used the M06-2X hybrid functional (Zhao & Truh-lar 2008) and the def2-TZVPP basis set (Weigend & Ahlrichs 2005; Weigend 2006) including D3 empirical dispersion corrections (Grimme et al. 2010) in Gaussian 09 (Frisch et al. 2009). All calculations were performed with tight convergence criteria and an ultrafine integration grid. Reactions were evaluated using relaxed scans—partial optimizations, where one coordinate corresponding to the forming bond was varied and all other coordinates were allowed to minimize. From these relaxed scans, minima were optimized and maxima were used as inputs for transition state searches. Transition states were characterized using intrinsic reaction coordinate scans to verify if these structures connected reactants and products. Harmonic vibrational analysis was carried out on the optimized structures; the resulting frequencies and zero-point energies were scaled by 0.94985 and 0.9748 respectively, as recommended by Kesharwani et al. (2015). Single point energies were refined by two CCSD(T) calculations using the augmented correlation consistent basis sets aug-cc-pVDZ and aug-cc-pVTZ, with the correlation energy ($CE$) extrapolated to the complete basis set limit ($CE_{\text{CBS}}$) using a modified version of the inverse power scheme taken from Helgaker et al. (1997) using the parameters found in Neese & Valeev (2011) and shown in equation 3. The self consistent field theory energies ($SCF$) were extrapolated using the exponential scheme taken from Zhong et al. (2008) using the parameters found in Neese & Valeev (2011) and shown in equation 4.

$$CE_{\text{CBS}} = (CE_{\text{ADZ}} \times 2^{2.46} - CE_{\text{ATZ}} \times 3^{2.46})/(2^{2.46} - 3^{2.46}) \tag{3}$$

$$SCF_{\text{CBS}} = SCF_{\text{ATZ}} + (e^{(4.42\sqrt{3}-4.42\sqrt{2})} - 1)^{-1} \times (SCF_{\text{ATZ}} - SCF_{\text{ADZ}}) \tag{4}$$

The reaction of SH with $c$-C$_3$H$_2$ has recently been explored by Flint & Fortenberry (2022), where the addition of SH to the ring followed by the loss of a ring-bound H atom to form $c$-C$_3$HSH was considered. These channels were found to be endothermic and have emerged barriers. In this work we considered additional channels: the loss of an H atom from the sulfonyl group, H atom transfers, and the elimination of H atoms from these intermediates and ring-opening reactions. Here, addition of SH to the C1 carbon (carbon with no H atoms bound to it, see top panel in Figure 2 for its structure) of the $c$-C$_3$H$_2$ ring to form $c$-C$_3$H$_2$SH-1 (TS 1a, -9.9 kJ mol$^{-1}$) is followed by a submerged barrier to elimination of an H atom from the sulfonyl group (TS 6, -26.6 kJ mol$^{-1}$) leading to the formation of $c$-C$_3$H$_2$S; this channel is exothermic by 29.3 kJ mol$^{-1}$ and is provided in the top panel of Figure 2. Additional submerged channels, following H atom transfer from the sulfonyl group to the ring and subsequent ring opening reactions can lead to the formation of both H$_2$CCCS and HCCCHS; a subset of the lowest energy pathways found are presented in the bottom panel of Figure 2.

The addition of SH to the $c$-C$_3$H$_2$ ring at the C2 position (the carbons with an H atom bound to them) leads to the formation of $c$-C$_3$H$_2$SH-2 which, following H atom elimination from the sulfonyl group, leads to the formation of HCCCHS. H atom transfer (TS 5) allows for the formation of $c$-C$_3$H$_2$SH-1 with the same channels to product. However, all these channels require the initial addition process to occur, and this appears to require surmounting an emerged barrier in the entrance channel (TS 1b, 28.8 kJ mol$^{-1}$); therefore, it is likely the predominant reaction from SH and $c$-C$_3$H$_2$ is via addition at C1.

This theoretical study highlights the potential of the reaction between SH and $c$-C$_3$H$_2$ to produce three isomers of C$_3$H$_2$S. Among these pathways, the formation of the cyclic isomer $c$-C$_3$H$_2$S is likely to be the preferred channel, due to the presence of a submerged barrier for H atom elimination following the initial addition step. Further theoretical and experimental work is required to evaluate the kinetics and product branching ratio of this reaction under conditions relevant to TMC-1.

## 6. CONCLUSIONS

We report the spectroscopic characterization of cyclopropenethione ($c$-C$_3$H$_2$S) in both laboratory experiments and astronomical observations, culminating in its detection using the GBT as part of the GOTHAM large program. This discovery marks the completion of the C$_3$H$_2$S isomeric family in TMC-1, as the other two isomers, CH$_2$CCS and HCCCHS were previously identified in the same source. Adding to the intriguing discoveries of sulfur-bearing species in TMC-1, the total column density of $c$-C$_3$H$_2$S is reported as $5.72^{+2.65}_{-1.61} \times 10^{10}$ cm$^{-2}$ at an excitation temperature of $4.7^{+1.3}_{-1.1}$ K — considerably lower than the reported column densities of both CH$_2$CCS and HCCCHS. This finding aligns with the RDP, a kinetic heuristic that predicts the most abundant isomer in a family to be the one with the smallest dipole moment assuming similar underlying chemistry. However, for the O-bearing molecules, the cyclic isomer exhibits a significantly *higher* column density than the linear isomer. As such, the linear form of the C$_3$H$_2$O isomer, which is highly reactive with atomic hydrogen in this region must react in fundamentally different way from its isovalent S-bearing analog. Overall, the RDP is now consistent between both the sulfur- and oxygen-bearing isomer families in TMC-1, although the reasons behind the behavior and the factors responsible for this remarkable difference in reactivity between the linear species remain unknown.

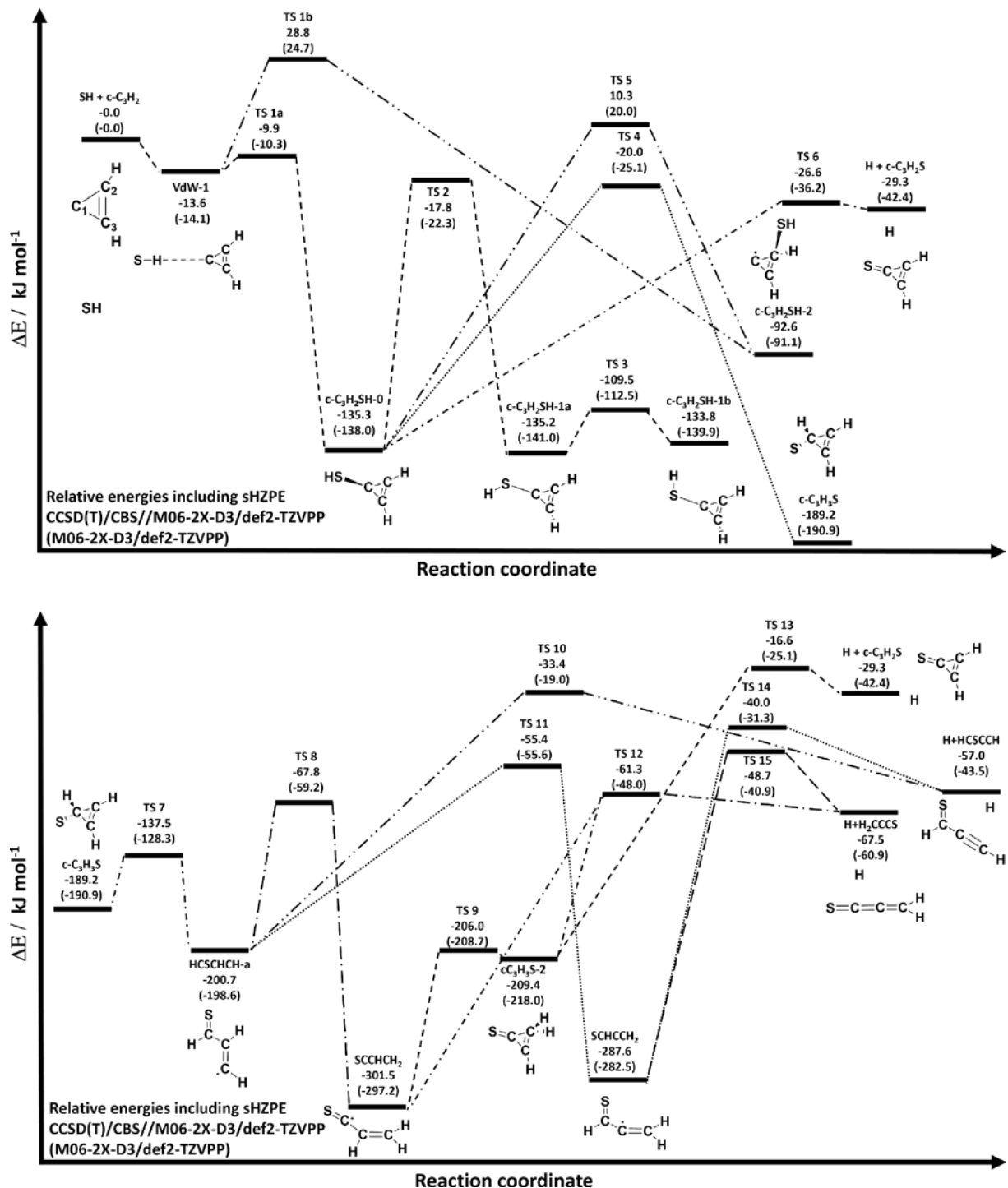

**Figure 2.** Potential energy surface for the reaction of SH with $c$-$C_3H_2$ at the CCSD(T)/CBS//M06-2X-D3/def2-TZVPP level. Energies are provided relative to the reactants in kJ mol$^{-1}$ including scaled harmonic zero point energy corrections (sHZPE). (*Left:*) Summary of the initial addition of SH to the ring and prompt formation of $c$-$C_3H_2S$. (*Right:*) Additional routes that can lead to the formation of $c$-$C_3H_2S$, HCCCHS and $CH_2CCS$.

## 7. DATA ACCESS & CODE

The code used to perform the analysis is part of the molsim open-source package (**?**).

*Facilities:* GBT

We gratefully acknowledge support from NSF grants AST-1908576 and AST-2205126. G.W. and B.A.M. acknowledge the support of the Arnold and Mabel Beckman Foundation Beckman Young Investigator Award. Z.T.P.F. and B.A.M. gratefully acknowledge the support of Schmidt Family Futures. I.R.C. acknowledges support from the Natural Sciences and Engineering Research Council of Canada (grant RGPIN-2022-04684), the Canada Foundation for Innovation and the B.C. Knowledge Development Fund. Support for H. N. S. and this work was provided by the NSF through the Grote Reber Fellowship Program administered by Associated Universities, Inc./National Radio Astronomy Observatory. The National Radio Astronomy Observatory is a facility of the National Science Foundation operated under cooperative agreement by Associated Universities, Inc. The Green Bank Observatory is a facility of the National Science Foundation operated under cooperative agreement by Associated Universities, Inc.

# APPENDIX

## A. MCMC FITTING RESULTS

We have fit $c$-C$_3$H$_2$O, $c$-C$_3$H$_2$S, CH$_2$CCS, CH$_2$CCO, and HCCCHS, using MCMC fits to the GOTHAM observations. Four emission features of $c$-C$_3$H$_2$O are well detected above the noise, as shown in Figure 3. The adopted prior probability distributions for each parameter are detailed in Table 8. The priors for $v_{LSR}$ of the four components are chosen to be a Gaussian distribution centered at approximately 5.6, 5.8, 5.9, and 6.0 km s$^1$, informed by previous GOTHAM analyses (Loomis et al. 2021). Meanwhile, all the remaining parameters have uninformative uniformly distributed priors. A corner plot of the posterior probability distributions and parameter covariances for the $c$-C$_3$H$_2$O fit is shown in Figure 4, with statistics of the marginalized posterior summarized in Table 9. We adopt the 50th percentile value of the posterior probability distributions as the representative value of each parameter to construct composite line profiles, as shown in Figure 3.

In order to achieve self-consistency, we adopted the same prior probability distributions for $c$-C$_3$H$_2$S and CH$_2$CCS, informed by the posterior distribution of $c$-C$_3$H$_2$O. The specific prior probability distributions for each parameter are detailed in Table 5. A corner plot of the parameter covariances and their distribution for the $c$-C$_3$H$_2$S MCMC fit is shown in Figure 5. An identical analysis to that for $c$-C$_3$H$_2$S was carried out for CH$_2$CCS, with a total of 36 transitions being considered for the MCMC ftting. The resulting posterior probability distributions for the CH$_2$CCS fit are shown in Figure 6, with statistics of the marginalized posterior summarized in Table 10. The majority of CH$_2$CCS emission is contributed by three of the four velocity components, a behavior similarly observed in other molecules towards TMC-1, such as propargyl cyanide and E-1-cyano-1,3-butadiene (McGuire et al. 2020; Cooke et al. 2023). There is not a significant detection in the component with a $v_{LSR}$ of 5.92 km s$^{-1}$. However, both four-component and three-component models result in similar total column densities. For consistency, we thereby present only the results assuming a four-velocity-component model.

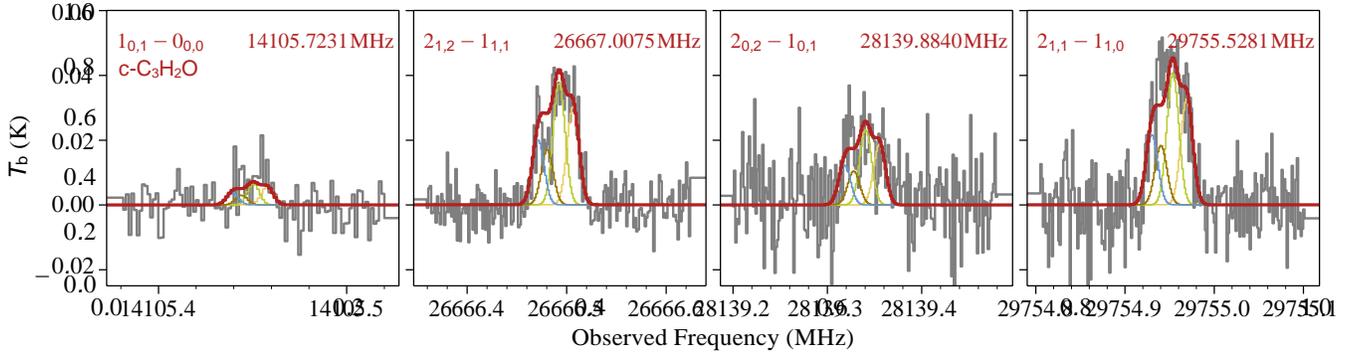

**Figure 3.** Individual line detections of $c$-C$_3$H$_2$O transitions in the GOTHAM DRV data. The observed spectra are displayed in black while the best-fitting model to the data, including all velocity components, is overlaid in red. Simulated spectra of the individual velocity components are shown in: yellow (5.62 km s$^{-1}$), lime (5.78 km s$^{-1}$), brown (5.92 km s$^{-1}$), and blue (6.02 km s$^{-1}$). See Table 9.

From the GOTHAM observations that were processed as part of DRV, no signal beyond a 1$\sigma$ detection limit can be assigned to CH$_2$CCO and HCCCHS. Column density upper limits are therefore constrained using the modified fitting process (Loomis et al. 2016). Source sizes, $T_{ex}$, $v_{LSR}$, and $\Delta V$ were held fixed to the 50th percentile value in $c$-C$_3$H$_2$O posteriors. Corner plots of column densities for CH$_2$CCO and HCCCHS are shown in Figure 7. Upper limits are given as the 97.8th percentile (2$\sigma$) value in Table 11.

**Table 8.** Priors used for MCMC analysis of $c$-$C_3H_2O$ observations towards TMC-1.

| Component No. | $v_{LSR}$ [km s$^{-1}$] | Size ["] | $\log_{10}(N_T)$ [cm$^{-2}$] | $T_{ex}$ [K] | $\Delta V$ [km s$^{-1}$] |
|---|---|---|---|---|---|
| 1 | $N(5.624, 0.01)$ | $U\{0, 300\}$ | | | |
| 2 | $N(5.790, 0.01)$ | $U\{0, 300\}$ | $U\{10.0, 12.0\}$ | $U\{2.73, 15.0\}$ | $U\{0.1, 0.3\}$ |
| 3 | $N(5.910, 0.01)$ | $U\{0, 300\}$ | | | |
| 4 | $N(6.033, 0.01)$ | $U\{0, 300\}$ | | | |

**Table 9.** Summary statistics of the marginalized $c$-$C_3H_2O$ posterior. Uncertainties of total $N_T$ are derived by adding the uncertainties of the individual components in quadrature.

| $v_{LSR}$ (km s$^{-1}$) | Size (") | $N_T$ ($10^{10}$cm$^{-2}$) | $T_{ex}$ (K) | $\Delta V$ (km s$^{-1}$) |
|---|---|---|---|---|
| $5.624^{+0.007}_{-0.007}$ | $183^{+79}_{-83}$ | $1.89^{+0.68}_{-0.26}$ | | |
| $5.781^{+0.008}_{-0.008}$ | $122^{+111}_{-73}$ | $2.42^{+1.40}_{-0.42}$ | $4.2^{+1.1}_{-0.7}$ | $0.158^{+0.015}_{-0.013}$ |
| $5.915^{+0.011}_{-0.010}$ | $132^{+110}_{-100}$ | $1.07^{+1.08}_{-0.40}$ | | |
| $6.018^{+0.009}_{-0.009}$ | $144^{+104}_{-95}$ | $1.26^{+0.70}_{-0.31}$ | | |
| $N_T$(Total): $6.64^{+2.02}_{-0.71} \times 10^{11}$ cm$^{-2}$ | | | | |

**Table 10.** Summary statistics of the marginalized $CH_2CCS$ posterior. Uncertainties of total $N_T$ are derived by adding the uncertainties of the individual components in quadrature.

| $v_{LSR}$ (km s$^{-1}$) | Size (") | $N_T$ ($10^{10}$cm$^{-2}$) | $T_{ex}$ (K) | $\Delta V$ (km s$^{-1}$) |
|---|---|---|---|---|
| $5.621^{+0.008}_{-0.008}$ | $185^{+78}_{-83}$ | $2.21^{+0.39}_{-0.34}$ | | |
| $5.780^{+0.009}_{-0.009}$ | $150^{+101}_{-96}$ | $1.78^{+0.57}_{-0.37}$ | $11.7^{+0.9}_{-0.8}$ | $0.160^{+0.020}_{-0.017}$ |
| $5.915^{+0.010}_{-0.010}$ | $128^{+116}_{-106}$ | $0.26^{+0.34}_{-0.12}$ | | |
| $6.020^{+0.008}_{-0.008}$ | $172^{+88}_{-91}$ | $1.55^{+0.37}_{-0.32}$ | | |
| $N_T$(Total): $5.80^{+0.85}_{-0.61} \times 10^{11}$ cm$^{-2}$ | | | | |

**Table 11.** Summary statistics of the marginalized $CH_2CCO$ and HCCCHS posterior.

| $v_{LSR}$ (km s$^{-1}$) | Size (") | $T_{ex}$ (K) | $\Delta V$ (km s$^{-1}$) | $CH_2CCO$ $N_T$ ($10^{11}$cm$^{-2}$) | HCCCHS $N_T$ ($10^{11}$cm$^{-2}$) |
|---|---|---|---|---|---|
| [5.624] | [183] | | | < 0.94 | < 2.69 |
| [5.781] | [122] | [4.2] | [0.158] | < 0.87 | < 2.13 |
| [5.915] | [132] | | | < 1.01 | < 2.10 |
| [6.018] | [144] | | | < 0.94 | < 2.59 |
| $CH_2CCO$ $N_T$(Total): $< 3.76 \times 10^{11}$ cm$^{-2}$ | | | | | |
| HCCCHS $N_T$(Total): $< 9.51 \times 10^{11}$ cm$^{-2}$ | | | | | |

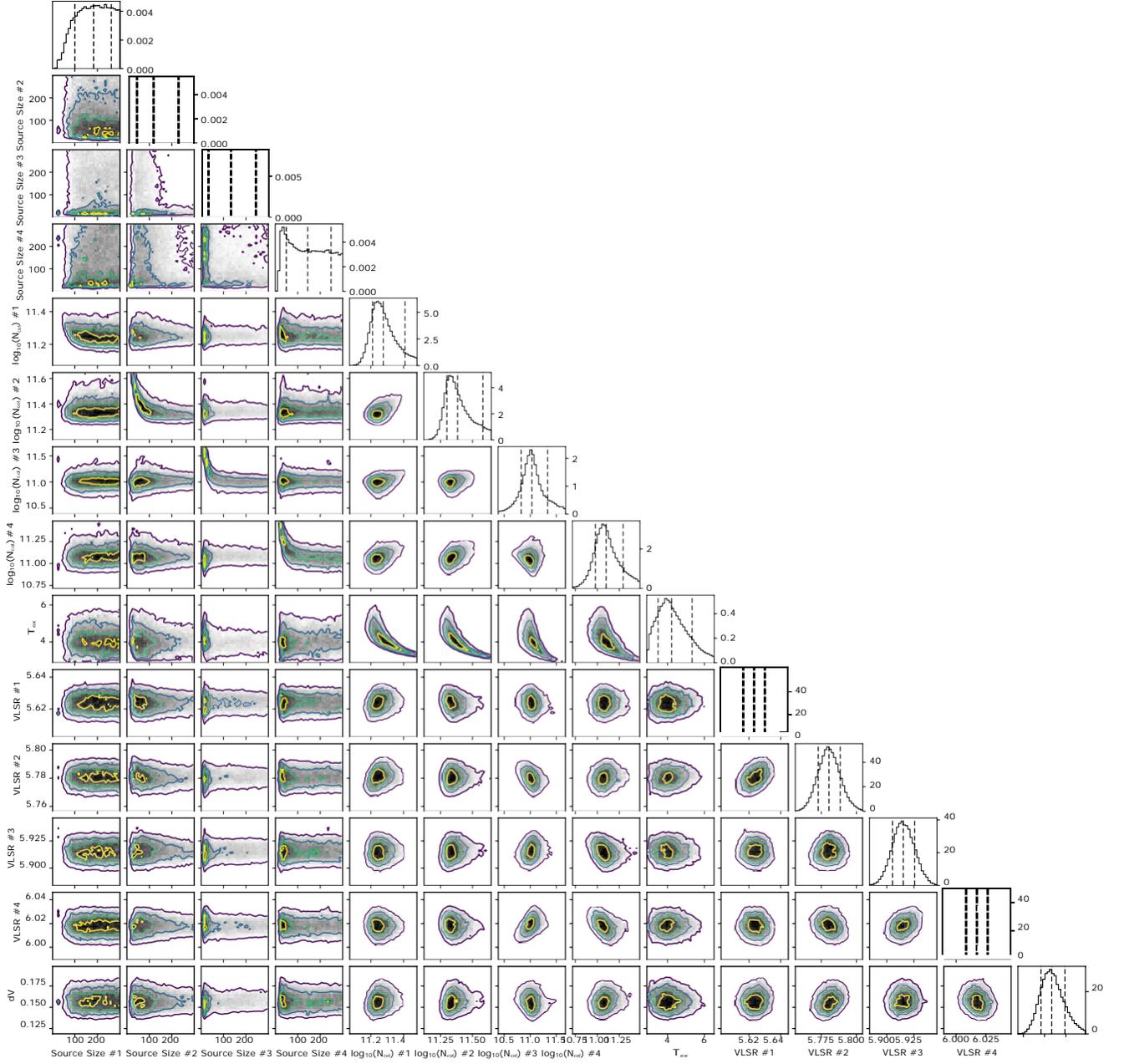

**Figure 4.** Parameter covariances and marginalized posterior distributions for the $c$-$C_3H_2O$ MCMC fit. $16^{th}$, $50^{th}$, and $84^{th}$ confidence intervals (corresponding to $\pm 1$ sigma for a Gaussian posterior distribution) are shown as vertical lines.

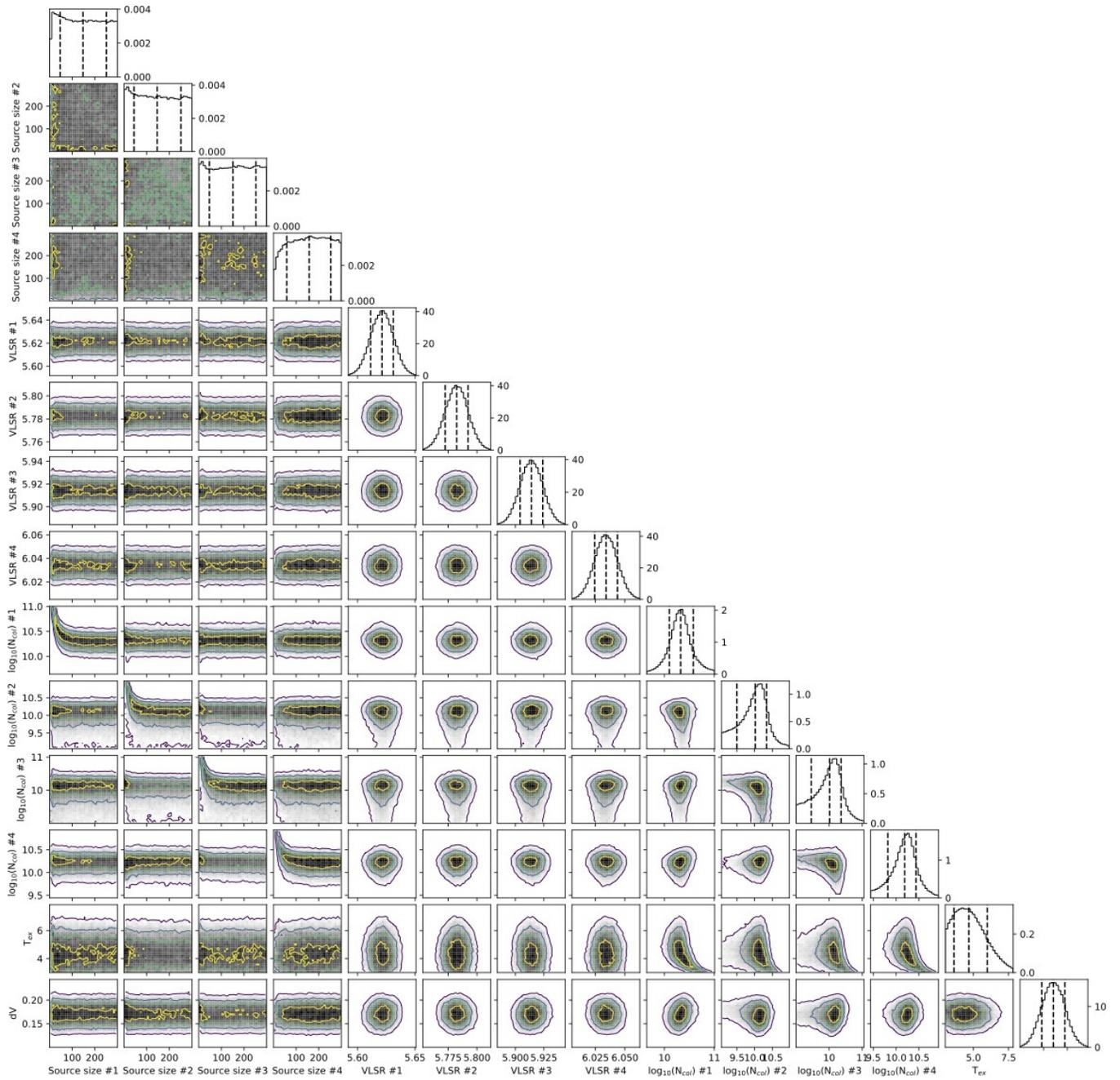

**Figure 5.** Parameter covariances and marginalized posterior distributions for the $c$-$C_3H_2S$ MCMC fit. $16^{th}$, $50^{th}$, and $84^{th}$ confidence intervals (corresponding to $\pm 1$ sigma for a Gaussian posterior distribution) are shown as vertical lines.

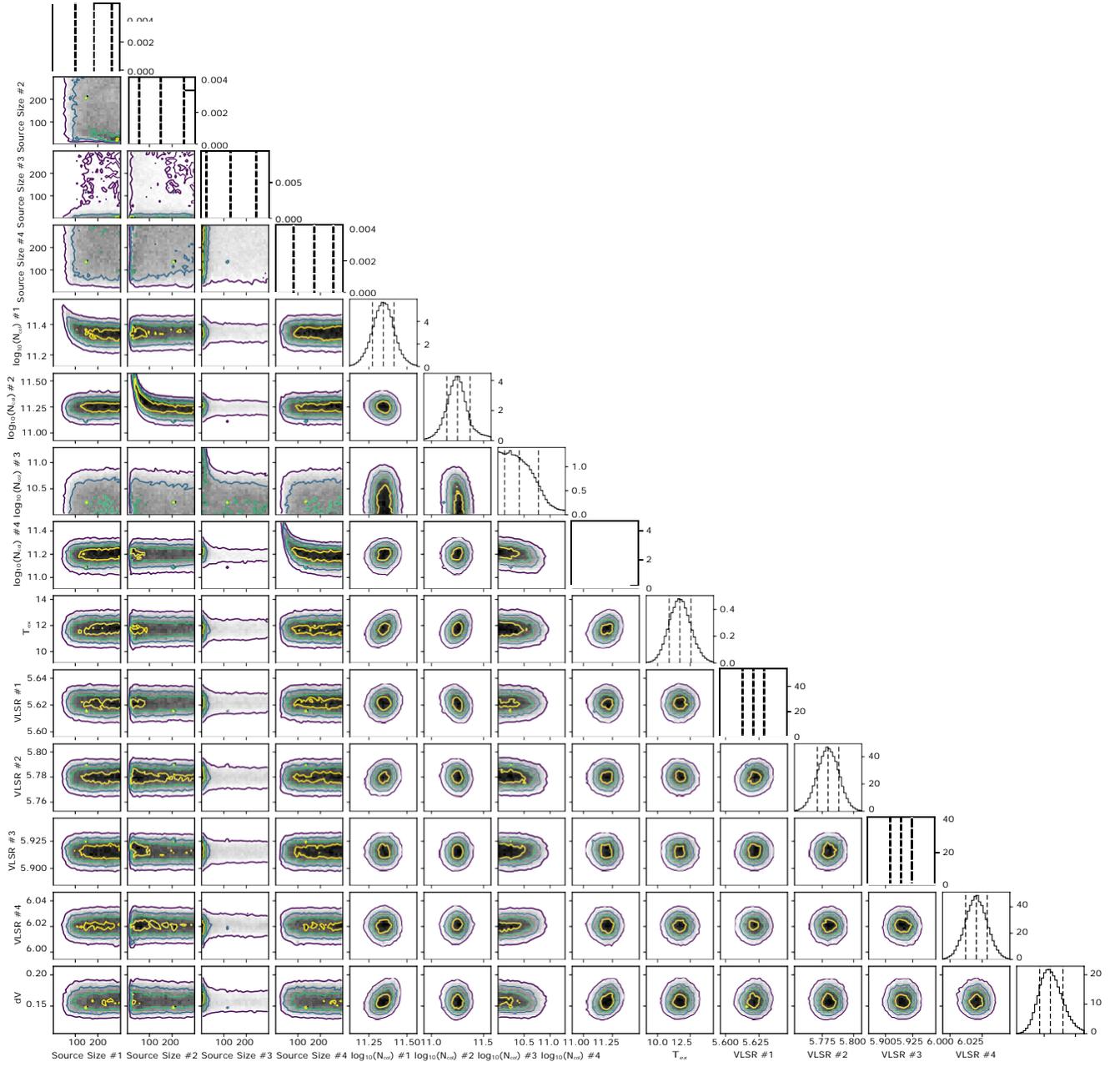

**Figure 6.** Parameter covariances and marginalized posterior distributions for the CH$_2$CCS MCMC fit. $16^{th}$, $50^{th}$, and $84^{th}$ confidence intervals (corresponding to $\pm 1$ sigma for a Gaussian posterior distribution) are shown as vertical lines.

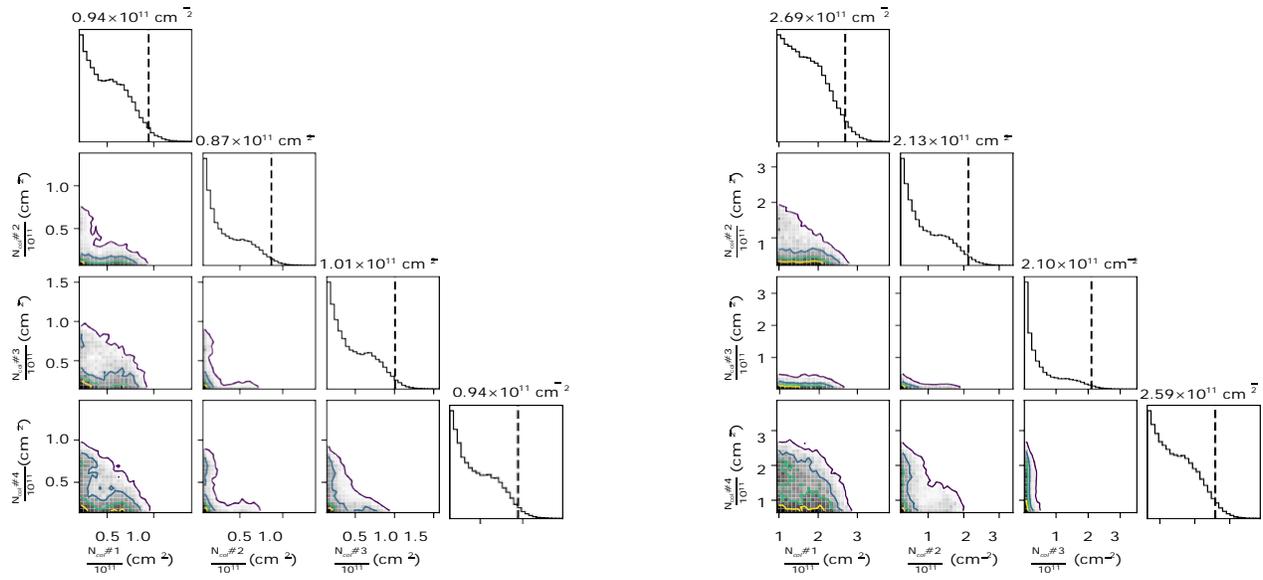

**Figure 7.** Marginalized posterior distributions of column densities for the H$_2$CCCO MCMC fit on the left panel and the HCCCHS MCMC fit on the right panel. 97.8$^{th}$ confidence intervals are shown as vertical lines.